\documentclass{aastex63}

\def\beq{\begin{equation}}
\def\eeq{\end{equation}}
\def\be{\begin{eqnarray}}
\def\ee{\end{eqnarray}}
\def\ba{\begin{eqnarray}}
\def\ea{\end{eqnarray}}

\def\L{\mathcal{L}}
\def\V{\mathcal{V}}
\def\D{\mathcal{D}}
\def\W{\mathcal{W}}
\def\M{\mathcal{M}}

\newcommand{\G}{\mathcal{G}}
\newcommand{\Z}{\mathcal{Z}}
\newcommand{\Y}{\mathcal{Y}}

\definecolor{darkred}{rgb}{.743,0,0}

\def\n02b{$0\nu\beta\beta$}
\def\n02bphi{$0\nu\beta\neta\phi$}
\def\lsim{\mathrel{\rlap{\lower4pt\hbox{\hskip1pt$\sim$}}
    \raise1pt\hbox{$<$}}}         
\def\gsim{\mathrel{\rlap{\lower4pt\hbox{\hskip1pt$\sim$}}
    \raise1pt\hbox{$>$}}}         

\begin{document}

\title{Could quasar lensing time delays hint to a core component in halos, instead of $H_0$ tension?}

\author{Kfir Blum}
\affiliation{Weizmann Institute, Department of Particle Physics and Astrophysics, Rehovot, Israel 7610001}
\affiliation{Theory department, CERN, 1 Esplanade des Particules, Geneva 23, CH-1211, Switzerland}
\author{Emanuele Castorina}
\affiliation{Theory department, CERN, 1 Esplanade des Particules, Geneva 23, CH-1211, Switzerland}
\author{Marko Simonovi\'c}
\affiliation{Theory department, CERN, 1 Esplanade des Particules, Geneva 23, CH-1211, Switzerland}

\email{kfir.blum@weizmann.ac.il}
\email{emanuele.castorina@cern.ch}
\email{marko.simonovic@cern.ch}

\begin{abstract}

The time delay measured between the images of gravitationally lensed quasars probes a combination of the angular diameter distance to the source-lens system and the mass density profile of the lens. Observational campaigns to measure such systems have reported a determination of the Hubble parameter $H_0$ that shows significant tension with independent determination based on the cosmic microwave background (CMB) and large scale structure (LSS). We show that lens mass models that exhibit a cored component, coexisting with a cusp, probe a degenerate direction in the lens model parameter space, being an approximate mass sheet transformation. This family of lens models has not been considered by the cosmographic analyses. Once added to the model, the cosmographic error budget should become dependent on stellar kinematics uncertainties. We propose that a core component coexisting with a cusp could bring the lensing measurements of $H_0$ to accord with the CMB/LSS value.

\end{abstract}


\section{Introduction and main result}
There appears to be a tension between measurements of the Hubble parameter $H_0$ based on the classic cosmic distance ladder method and measurements obtained through a fit of the standard $\Lambda$CDM model to the CMB or LSS data (for a summary, see \cite{Verde19}). The SH0ES collaboration \citep{SHOES} reported $H_0 = 74.03\pm 1.42$ km/s/Mpc using supernovae calibrated with Cepheids. This result is more than 4$\sigma$ discrepant with the best fit $\Lambda$CDM value given by the Planck collaboration $H_0 = 67.36 \pm 0.54$ km/s/Mpc~(\cite{Planck18}), or with $H_0$ measurements obtained from galaxy clustering and galaxy lensing data \citep{DESH0,Ivanov19,DAmico,Troster19}, that are independent of but agree with the CMB result. The tension is not so strong in the analysis of a supernova sample calibrated with the tip of the red giant branch, that gives $H_0=69.8\pm 0.8\text{(stat)}\pm1.7(\text{sys})$ km/s/Mpc~(\cite{Freedman19}). 
On the other hand, a third, alternative method to constrain $H_0$ independently of both the distance ladder and cosmological perturbation theory, is provided by measurements of time delays in strongly lensed systems~(\cite{Refsdal64,Kochanek_2002,Kochanek06}).
The H0LiCOW collaboration~(\cite{Suyu17}) used the time delays between multiple images of strongly lensed galaxies hosting a quasar to obtain $H_0=73.3^{+1.7}_{-1.8}$ km/s/Mpc \citep{Suyu17,Bonvin19,Birrer19,Chen19,Wong19}, achieving 2.5\% precision with a central value in agreement with SH0ES. Combining the SH0ES and H0LiCOW measurements, the result is in more than $5\sigma$ tension with CMB/LSS data.

Given the importance of these results to cosmology, it is worth investigating them from every angle. In this paper we focus on the lensing time delay (``cosmography") measurements. It is well known that lensing analyses are subject to systematic uncertainty associated with the choice of the family of models used to reconstruct the lens potential~(\cite{Falco1985,SchneiderSluse13A,SchneiderSluse13B,Xu16,Unruh16,Tagore17,Sonnenfeld18,GomerWilliams19,Kochanek19}). The H0LiCOW collaboration is, of course, well aware of this problem, and had taken measures to mitigate it by considering different families of lens models. Nevertheless, H0LiCOW systems probe the baryonic-dominated inner part of the lens, where theoretical understanding of the mass profile is limited to challenging hydrodynamical simulations. Moreover, given that we do not know what makes up the dark matter, its distribution on galactic scales could exhibit unexpected features. With these issues in mind, the question we address in this paper is: could a feature in the mass density profile of H0LiCOW lenses bring the cosmographic result for $H_0$ to agree with the CMB/LSS value? 
Another way to phrase this question could be to accept, tentatively and for the purpose of the exercise, the CMB/LSS value of $H_0$; and then ask, given this hypothesis, what would the cosmography data teach us about the inner structure of galaxies.

We believe that we have found an interesting answer to this question. To summarise, we find that if one took the simple power law (PL) density models, shown by H0LiCOW to provide a good fit to the lensing data, and then added a core component {\it in addition to} the PL, then: (i) the lensing reconstruction problem should be equally well solved by the PL+core models as it is for the pure PL; and (ii) the addition of comparable cores to all H0LiCOW lenses would systematically shift the inferred cosmographic value of $H_0$ downwards in all systems, in accord with the fact that all of the systems analysed by the standard H0LiCOW analysis pipeline consistently hint to high $H_0$. 

The cores we need are a moderate deformation of the nominal profile: at the Einstein radius (translating to a few kpc for H0LiCOW systems, where the lenses are massive elliptical galaxies), the core component need only make-up 10\% or less of the total enclosed mass of the lens. Outside of the Einstein radius the relative core contribution could become larger, potentially reaching as much as $\mathcal{O}$(1) of the mass and opening a possible way to constrain our solution with detailed kinematic modelling. However, current lensing data do not constrain very well the outer extent of the core. 

Single-source lensing data cannot distinguish a pure PL profile from PL+core, because moving along the PL+core family of models (as we shall define it in Sec.~\ref{Sec:lPL}) is an approximate mass sheet transformation (MST)~(\cite{Falco1985}). Therefore, PL+core models probe a flat direction in the likelihood for $H_0$.  
Stellar kinematics could, in principle, break the mass sheet degeneracy (MSD) \citep{Romanowsky_1999,TreuKoopmans02,Jee_2015,Jee_2016,ShajibKin}. However, as mentioned above, to solve the $H_0$ tension we need an effect of no more than 10\% in enclosed mass within $\theta_E$. Constraining this with stellar kinematics would not be trivial and would suffer from systematic uncertainties related to, e.g., the velocity anisotropy modelling. Furthermore, kinematics modelling uncertainties would come to dominate the determination of $H_0$, which should be revised~(\cite{Kochanek19}). 
Perhaps another potential way to resolve the MSD would be to have multiple sources lensed by the same object, as is usually the case in lensing by galaxy clusters \citep{Grillo18,Grillo19}. 

It is worth pointing out that we are not aware of the presence of PL+core profiles in simulations. In this sense, introducing them is an ad-hoc solution of the (cosmographic contribution to) the $H_0$ tension. But we are also not aware of observational data that excludes such profiles. If one accepts PL+core profiles, 
then the question arises what is the core component made of. Since the inner part of the lenses is baryon-dominated, we do not know at this point if the core is some baryonic structure, or dark matter. It is exciting to speculate that the $H_0$ tension could actually hint to new constraints on the nature of the dark matter, perhaps along the lines of models such as in \cite{Schive:2014dra} or \cite{Spergel:1999mh}.

This paper is organized as follows. In Sec.~\ref{Sec:lPL} we show that adding an inner core component, on top of a (rescaled) cusp component, is an approximate MST. We give some simple examples, estimate the MSD breaking effects, and introduce $\lambda$PL models as a family of models that is expected to probe a flat direction in cosmographic measurements of $H_0$. In Sec.~\ref{Sec:lenses} we consider as input the CMB/LSS measurement of $H_0$ and use it to estimate the required morphology of H0LiCOW lenses. In Sec.~\ref{Sec:disc} we discuss our findings, and the possibility that the (cosmographic part of the) $H_0$ tension might actually hint to a core component coexisting with a central cusp in galaxies. In App.~\ref{app:1} we collect some formulae for profiles that could serve as $\lambda$PL models.

\section{Adding a core to a cusp is a mass sheet transformation}
\label{Sec:lPL} 
Lensing analyses \citep{Suyu17,Bonvin19,Birrer19,Chen19,Wong19} take as input a brightness map defined on the image plane, spanned by coordinates $\vec\theta$, and constrain the deflection angle $\vec\alpha(\vec\theta)$ given by
\be\label{eq:ams}\vec\alpha(\vec\theta)&=&\frac{1}{\pi}\int d^2\theta'\frac{(\vec\theta-\vec\theta')}{|\vec\theta-\vec\theta'|^2}\kappa(\vec\theta')
\ee
via solving the lens equation
\be
\label{eq:bms}\vec\beta&=&\vec\theta-\vec\alpha(\vec\theta),\ee
where $\vec\beta$ parametrizes positions on the source plane.
The deflection angle is obtained by integration over the convergence,
\be\label{eq:kap}\kappa(\vec\theta)&=&\frac{\Sigma(\vec\theta)}{\Sigma_c},\\
\label{eq:Sc}\Sigma_c&=&\frac{D_s}{4\pi GD_lD_{ls}}.
\ee
Here $\Sigma(\vec\theta)$ is the projected surface mass density of the lens and $D_s,\,D_l,\,D_{ls}$ are the angular diameter distances from the source to the observer, from the lens to the observer, and from the source to the lens. 

Given multiple images of a quasar contained in the host galaxy, one  constructs the time delay $\Delta t_{ij}$ between quasar images $\vec\theta_i$ and $\vec\theta_j$,
\be\label{eq:Dt}\Delta t_{ij}&=&\mathcal{D}\Delta\tau_{ij},\\
\label{eq:Dtau}\Delta\tau_{ij}&=&\frac{\vec\alpha^2(\vec\theta_i)-\vec\alpha^2(\vec\theta_j)}{2}+\psi(\vec\theta_j)-\psi(\vec\theta_i),\\
\label{eq:D}\mathcal{D}&=&(1+z_l)\frac{D_sD_l}{D_{ls}},\ee
where
\be\label{eq:psims}\psi(\vec\theta)&=&\frac{1}{\pi}\int d^2\theta'\kappa(\vec\theta')\ln|\vec\theta-\vec\theta'|.\ee
If one has a model of $\kappa(\vec\theta)$, then it can be used to calculate $\Delta\tau_{ij}$ in Eq.~(\ref{eq:Dtau}). Given a measurement of $\Delta t_{ij}$, one can extract $\mathcal{D}=\Delta t_{ij}/\Delta\tau_{ij}$ and thus $H_0\propto1/\mathcal{D}\propto\Delta \tau_{ij}/\Delta t_{ij}$.

The MSD comes from the fact that if the lensing reconstruction problem Eq.~(\ref{eq:bms}) is solved by a model for $\kappa(\vec\theta)$, along with a model for the source position $\vec\beta$, then the reconstruction problem is also solved equally well by the alternative MST model\footnote{We note that the MST represents a subset of a more general set of transformations that leave the lens equation invariant~(\cite{SchneiderSluse13B}).} 
\be\label{eq:kl}\kappa_\lambda(\vec\theta)&=&\lambda\kappa(\vec\theta)+1-\lambda,\\
\vec\beta_\lambda&=&\lambda\vec\beta,
\ee
leaving $\vec \theta$ unchanged. 
While the lensing image plane geometry is invariant under the MST, the time delay is not invariant and it is easy to verify that $\Delta\tau_{ij,\lambda}=\lambda\Delta\tau_{ij}$. This means that if we measure $H_0$ from some model $\kappa$, then the MST model $\kappa_\lambda$ would give $H_0\to\left(\Delta\tau_{ij,\lambda}/\Delta\tau_{ij}\right)H_0=\lambda H_0$. 

The actual reconstruction problem of H0LiCOW deals with an extended source model, given by a map of the brightness $I_s(\vec\beta)$ on multiple source plane pixels. Under an MST, the distortion matrix $A_{ij} (\vec \theta) = \partial \beta_i /\partial \theta_j$ is rescaled to $A_{\lambda, ij} (\vec \theta) = \lambda A_{ij} (\vec \theta)$, which means that the magnification $\mu = 1/ \mathrm{det}A$ becomes $\mu_\lambda  =\mu /\lambda^2$. Importantly, the relative magnification between images is unchanged. For H0LiCOW systems the precise intrinsic luminosity of the source is unknown, so absolute magnification cannot be measured. Thus extended source information does not mitigate the MSD. 

The ``mass sheet" in MSD refers to the $1-\lambda$ term in Eq.~(\ref{eq:kl}) which is a constant convergence term and thus it acts as a $\vec\theta$-independent mass sheet. H0LiCOW took careful measures to account for the MSD due to external convergence $\kappa_{\rm ext}$ (see dedicated discussions in \citep{Suyu17,Bonvin19,Birrer19,Chen19,Wong19}). This was done by using numerical simulations to estimate the cumulative contributions of mass along the line of sight in the field of the lens systems. 
However, a cored density profile (3D density $\rho\sim$~const) extending over a finite radius $R_c$, and dropping quickly afterwards, can also give $\kappa$ that is constant inside\footnote{Here and elsewhere $\theta=|\vec\theta|$.} $\theta\lesssim \theta_c=R_c/D_l$. Thus, if the images in the lensing data only extend over angles $\theta<\theta_c$, then the addition of a cored density component, with $\kappa_c$ that is constant inside $\theta_c$, is (i) an approximate MST, if it is done alongside a rescaling of the previous $\kappa$ model, and (ii) is not equivalent to an external convergence term\footnote{See~\cite{SchneiderSluse13A} for a closely related discussion.}.

To make things more concrete we define the $\lambda$PL family of profiles:
\be\label{eq:PLlam}\kappa_\lambda(\vec\theta)&=&\lambda\kappa_{\rm PL}(\vec\theta)+(1-\lambda)\kappa_{\rm c}(\vec\theta).
\ee
Here, we take $\kappa_{\rm PL}$ to represent the elliptic PL profile as used by H0LiCOW to successfully model the lensing data in their systems\footnote{The elliptic PL profile is referred to as SPEMD in \citep{Suyu17,Bonvin19,Birrer19,Chen19,Wong19}.}.
The $\kappa_c(\vec\theta)$ term is chosen to satisfy $\kappa_c(\vec\theta)\approx1$ for $\theta<\theta_c$ and to fall faster than $\kappa_{\rm PL}$ at $\theta>\theta_c$. We do not need to assume that $\kappa_c(\vec\theta)$ is isotropic, but in what follows for simplicity we will. 

As a first example, consider the 3D cored density profile $\rho_c(r)=\frac{2}{\pi}\Sigma_cR_c^3(R_c^2+r^2)^{-2}$, where $\Sigma_c$ is the critical density of Eq.~(\ref{eq:Sc}). The convergence for this profile is $\kappa_c(\vec\theta)=\left(1+\frac{\theta^2}{\theta_c^2}\right)^{-\frac{3}{2}}=1-\frac{3\theta^2}{2\theta_c^2}+\mathcal{O}\left(\frac{\theta^4}{\theta_c^4}\right)$ and it induces the deflection angle $\vec\alpha_c(\vec\theta)=\hat\theta\frac{2\theta_c^2}{\theta}\left(1-\left(1+\frac{\theta^2}{\theta_c^2}\right)^{-\frac{1}{2}}\right)=\vec\theta\left(1-\frac{3\theta^2}{4\theta_c^2}+\mathcal{O}\left(\frac{\theta^4}{\theta_c^4}\right)\right)$. Obviously, using this $\kappa_c$ in Eq.~(\ref{eq:PLlam}) gives an approximate MSD inside of $\theta<\theta_c$. We can estimate the corrections to the MSD by comparing the Einstein angle $\theta_E$ for $\kappa_{\rm PL}$ and the Einstein angle $\theta_{E\lambda}$ for $\kappa_\lambda$ in Eq.~(\ref{eq:PLlam}). For simplicity, in this exercise we take $\kappa_{\rm PL}$ to be isotropic and given by $\kappa_{\rm PL}(\vec\theta)=\frac{3-\gamma}{2}\frac{\theta_E^{\gamma-1}}{\theta^{\gamma-1}}$, for which the deflection angle is $\vec\alpha_{\rm PL}(\vec\theta)=\frac{\theta_E^{\gamma-1}}{\theta^{\gamma-1}}\vec\theta$. 
In the limit $\theta_c\to\infty$, the MSD is exact and $\theta_E=\theta_{E\lambda}$. For finite $\theta_c$ we find $\theta_{E\lambda}=\theta_E+\delta$, with $\delta=-\frac{3}{4(\gamma-1)}\frac{1-\lambda}{\lambda}\frac{\theta^2_E}{\theta^2_c}+\mathcal{O}\left(\frac{\theta_E^4}{\theta_c^4}\right)$. From the form of $\delta$ we can infer the parametric dependence of the breaking of the MSD. The corrections to the image plane geometry enter at order $\theta^2/\theta_c^2$, and if $\lambda\approx1$ (that is, if we only add a small core) are further suppressed by a factor $1-\lambda$. Note that for real systems H0LiCOW find $\gamma\approx2$  so $1/(\gamma-1)\approx1$ (see Tab.~\ref{Tab:lenses}). 

More generally, if in Eq.~(\ref{eq:PLlam}) we use a core component that can be expanded as $\kappa_c=1+a\theta^2/\theta_c^2+...$ at $\theta<\theta_c$, then the leading order image plane corrections to the MSD at $\theta<\theta_c$ scale as $(1-\lambda)\theta^2/\theta_c^2$. This scaling remains true also when the baseline term $\kappa_{\rm PL}$ (or whatever other baseline model is considered, e.g.~a composite stellar cusp+NFW model) is anisotropic. 

\begin{figure}
\centering
 \includegraphics[scale=0.5]{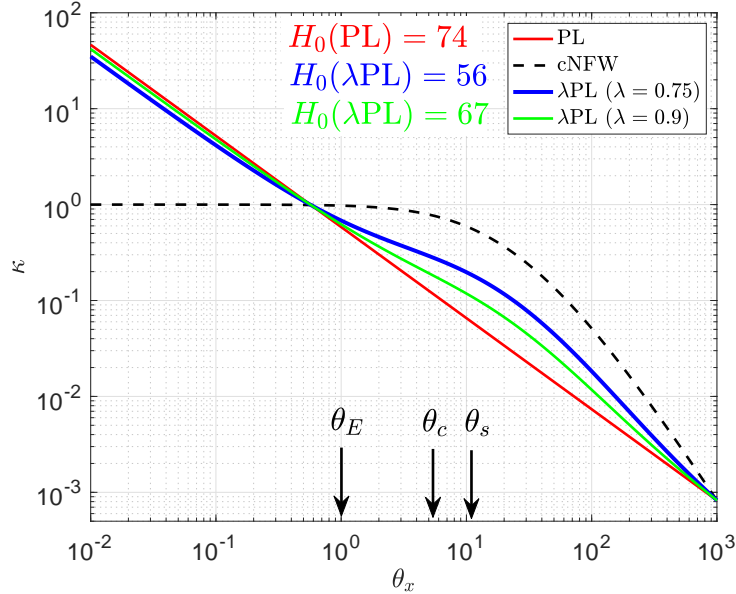}
 \caption{Convergence for a $\lambda$PL model, with $\lambda=0.75$ (blue) and $\lambda=0.9$ (green). The $\lambda=1$ pure PL case is shown in red, and the cNFW profile is shown in dashed black. A value of $\lambda\approx0.9$ would bring the H0LiCOW determination of $H_0$ down to the CMB/LSS value. Note that for H0LiCOW lenses, both lensing and kinematics data reach outward only slightly beyond $\theta_E$, and never constrain angles around the value of $\theta_c$ chosen in this example.}
 \label{fig:examplekappa}
\end{figure}

As another example, consider the 3D cored Navarro-Frenk-White (NFW)  density profile,
\be\rho_{\rm cNFW}(r)&=&\frac{\rho_0}{(R_c+r)(R_s+r)^2},\ee
which contains 1 extra parameter $R_c$, defining the core, in addition to the usual NFW density $\rho_0$ and scale radius $R_s$. The convergence $\kappa_{\rm cNFW}$ can be computed analytically even though is not particularly illuminating (in App.~\ref{app:1} we collect some formulae for profiles that could serve as the core component in $\lambda$PL models). With proper normalization such that $\kappa_{\rm cNFW}(0)=1$ it has the correct characteristics to function as $\kappa_c$ in Eq.~(\ref{eq:PLlam}).
We show $\kappa_{\rm cNFW}$ by the dashed black line in Fig.~\ref{fig:examplekappa}. We have set $\theta_s=R_s/D_l=11$ and $\theta_c=0.5\, \theta_s$, indicated by arrows at the bottom of the plot.

\begin{figure}
\centering
 \includegraphics[scale=0.6]{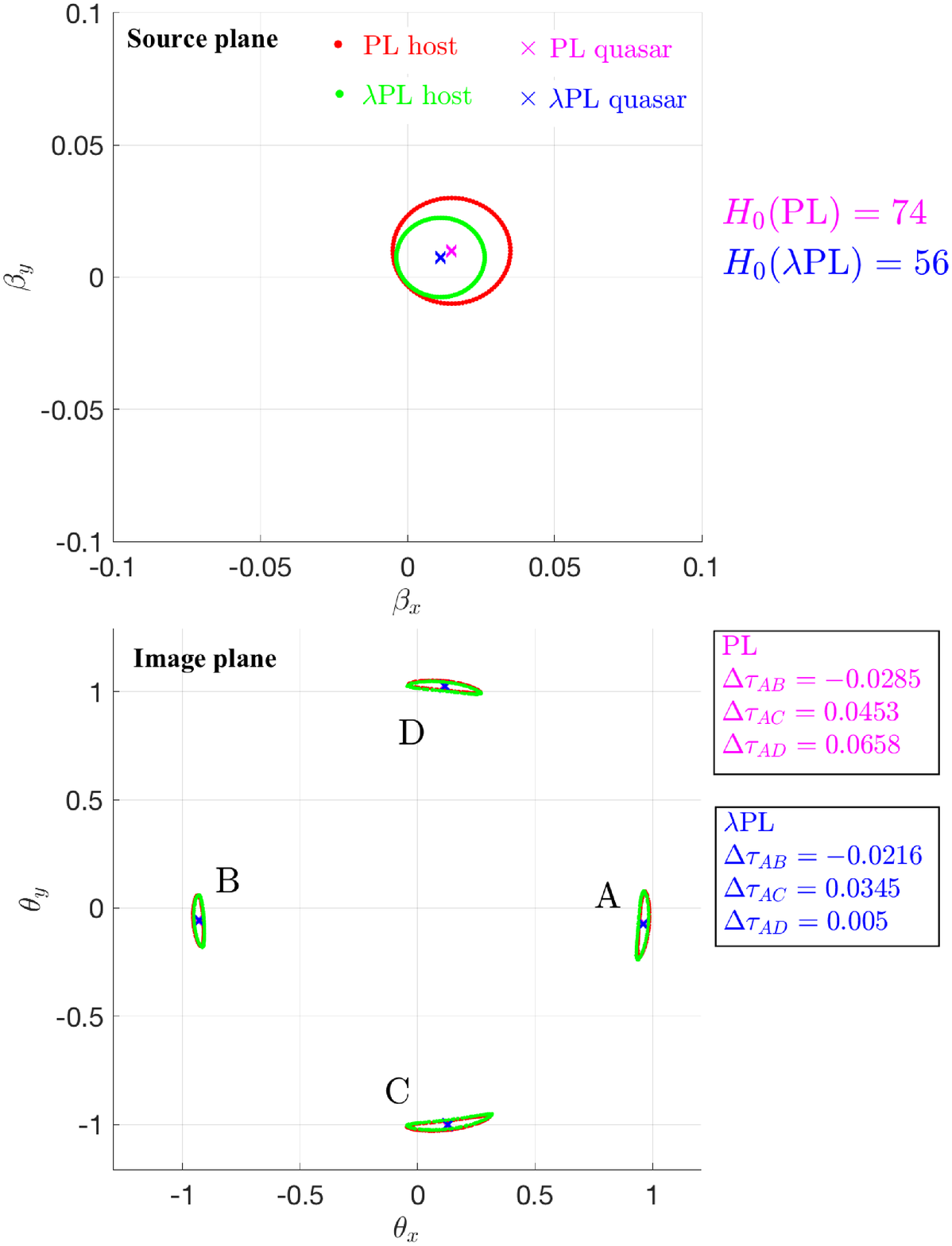}
 \caption{Example of PL-core MST for $\lambda=0.75$. {\bf Top:} Source plane. {\bf Bottom:} Image plane. Dimensionless time delays in inset.}
 \label{fig:example}
\end{figure}

To illustrate the MSD, in Fig.~\ref{fig:example} we calculate the lensing geometry and time delays for a toy model of a quasar sitting in an extended host galaxy. To make things simple we replace the extended host by a circle on the source plane, centred on the quasar. We first do the lensing exercise for a pure PL model with slope $n=1.95$ and ellipticity parameter $q=0.8$, similar to typical H0LiCOW systems. The source plane host ``galaxy" is shown by the red circle in the top panel (source plane). The ``quasar" is denoted by magenta cross. The lensed images are shown by red lines in the bottom panel (image plane). (It is difficult to see these lines because they lie underneath the green lines of the $\lambda$PL model, as explained below.) We calculate the dimensionless time delays $\Delta\tau_{ij}$ at the quasar image positions and show them next to the bottom panel (magenta, titled PL). The convergence for this PL model (along the $\theta_x$ axis) is shown by the red line in Fig.~\ref{fig:examplekappa}. We have chosen the PL normalisation such that $\theta_E\approx1$. 

Next, we consider a $\lambda$PL model with $\lambda=0.75$. The convergence for this $\lambda$PL model is shown by the blue line in Fig.~\ref{fig:examplekappa}. The source plane host model as given by the MSD is shown by the green circle in the top panel of Fig.~\ref{fig:example}. The quasar is shown by the blue cross. The images are shown by the green line and blue crosses in the bottom panel. As expected, they sit {\it almost} on top of the pure PL. 
The time delays for the $\lambda$PL model images are shown next to the bottom panel (blue, titled $\lambda$PL). As expected the $\lambda$PL time delays satisfy $\Delta\tau_{ij,\lambda}\approx\lambda\Delta\tau_{ij}$.
%

\section{If we assume $H_0$ from CMB/LSS, what do we learn about H0LiCOW lenses?}\label{Sec:lenses}
If one used the toy example of Fig.~\ref{fig:example} to measure $H_0$, and if, assuming the pure PL model, one found, for example, $H_0=74$~km/s/Mpc, then we expect that the $\lambda$PL model with $\lambda=0.75$ would give acceptable likelihood with $H_0\approx56$~km/s/Mpc.
Our choice of $\lambda$ in this example is, of course, an exaggeration. To solve the $H_0$ tension we only need $\lambda\approx0.9$. 
In Tab.~\ref{Tab:lenses} we collect some key numbers for six H0LiCOW systems. Taking $H_0\approx67$~km/s/Mpc to represent the CMB/LSS measurement, we show in the third column the value of $\lambda$ that is required to bring the cosmographic $H_0$ from each system down to the CMB/LSS value.
\begin{table}[htp]
\caption{Lens systems from \cite{TDCOSMO}. Values for $H_0$ (in km/s/Mpc) are from the PL fit  (Fig.~6 in \cite{TDCOSMO}). Approximate values for the PL index $\gamma$, the Einstein radius $\theta_E$, and the NFW scale $\theta_s$ were read from PL and composite NFW+stellar fits reported by papers in the last column.}
\begin{center}
\begin{tabular}{|c|c|c|c|c|c|c|c|}
\hline
&$H_0$&$\lambda=67/H_0$&$\gamma$&$\theta_E$~['']&$\theta_s$~['']&lens redshift $z_l$&ref\\
\hline\hline
RXJ1131&$76.1^{+3.6}_{-4.3}$&$0.88^{+0.06}_{-0.04}$&$1.98$&$1.6$&$19$ &$0.295$&\cite{Chen16}\\
\hline
PG1115&$83.0^{+7.8}_{-7.0}$&$0.81^{+0.07}_{-0.07}$&$2.18$&$1.1$&$17$ &$0.311$&\cite{Chen19}\\
\hline
HE0435&$71.7^{+5.1}_{-4.6}$&$0.93^{+0.07}_{-0.06}$&$1.87$&$1.2$&$10$&$0.4546$&\cite{Chen19}\\
\hline
DESJ0408&$74.6^{+2.5}_{-2.9}$&$0.9^{+0.03}_{-0.03}$&$2$&$1.9$&$13$&$0.6$&\cite{Shajib19}\\
\hline
WFI2033&$72.6^{+3.3}_{-3.5}$&$0.92^{+0.05}_{-0.04}$&$1.95$&$0.9$&$11$ &$0.6575$&\cite{Rusu19}\\
\hline
J1206&$67.0^{+5.7}_{-4.8}$&$1^{+0.08}_{-0.08}$&$1.95$&$1.2$ &4.7&$0.745$&\cite{Birrer19}\\
\hline
\end{tabular}
\end{center}
\label{Tab:lenses}
\end{table}%

\begin{figure}
\centering
 \includegraphics[scale=0.45]{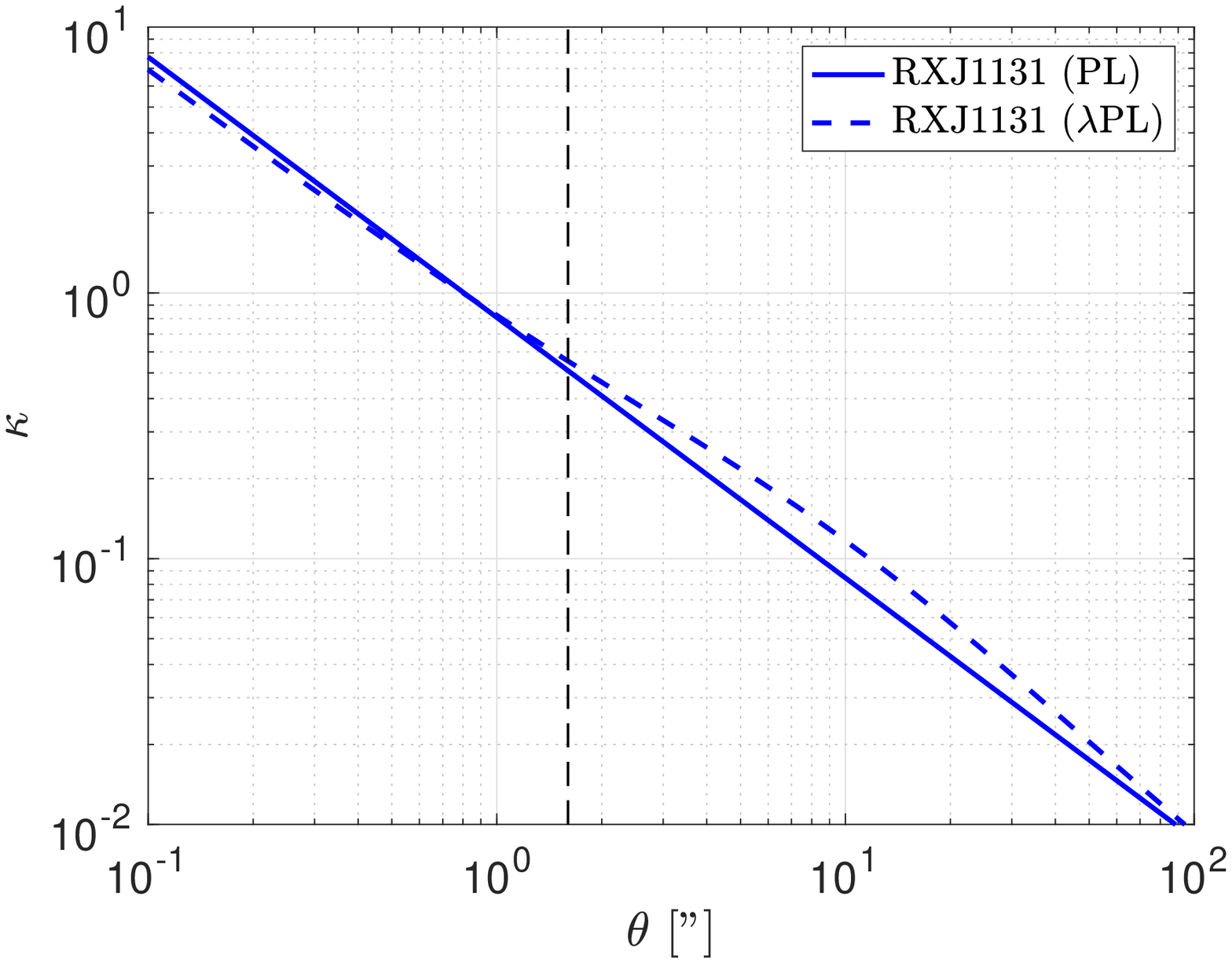}
 \includegraphics[scale=0.45]{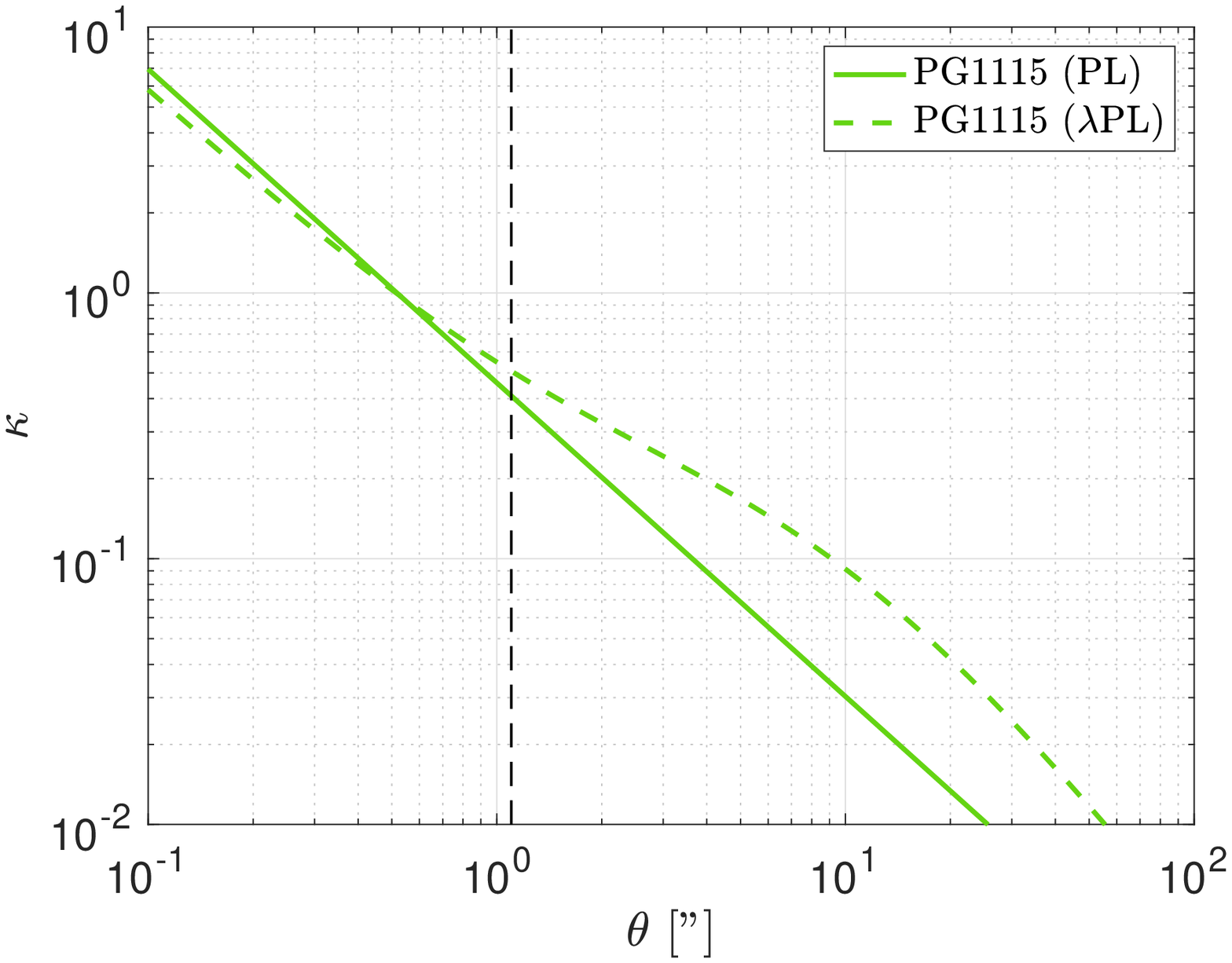}
  \includegraphics[scale=0.45]{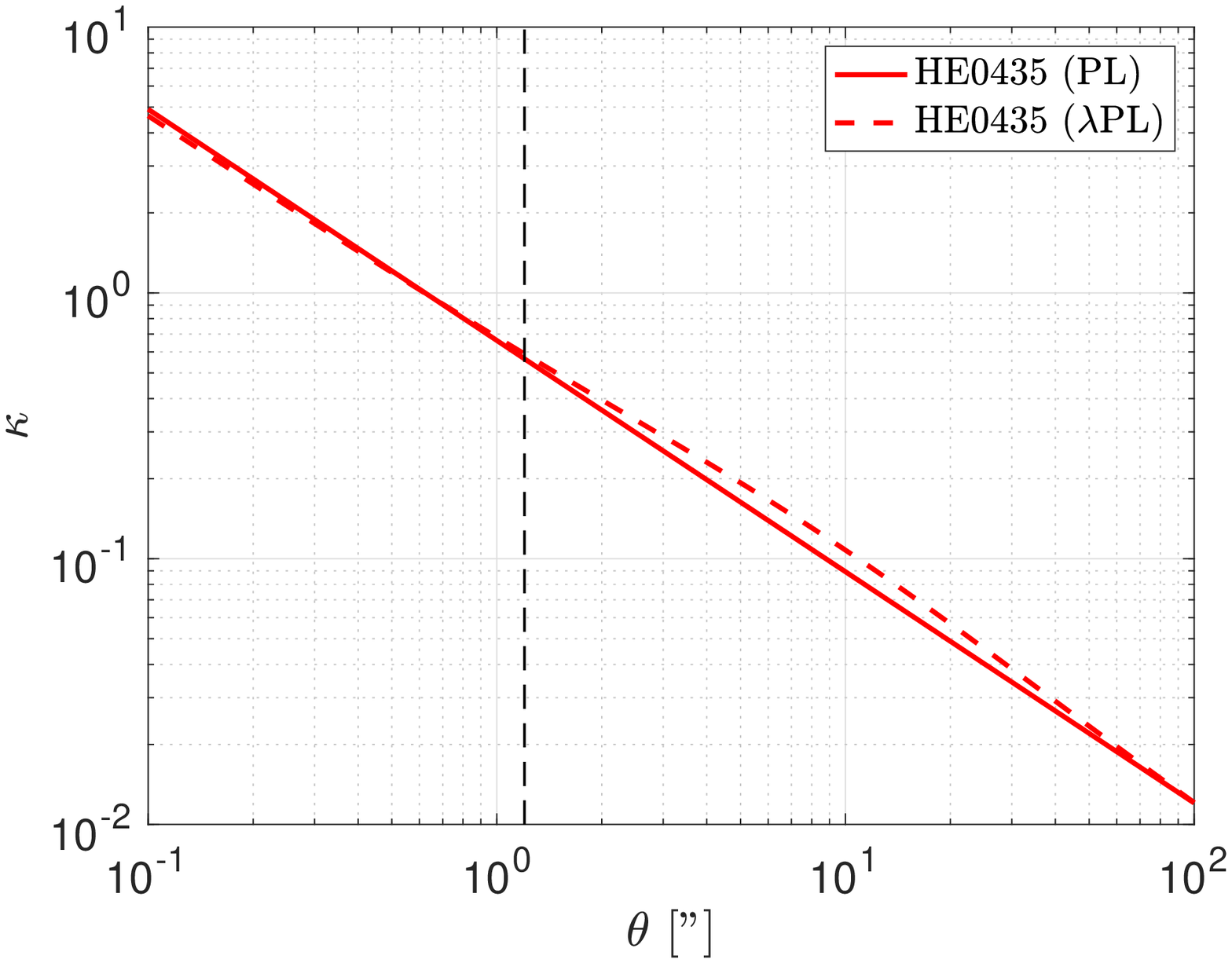}
   \includegraphics[scale=0.45]{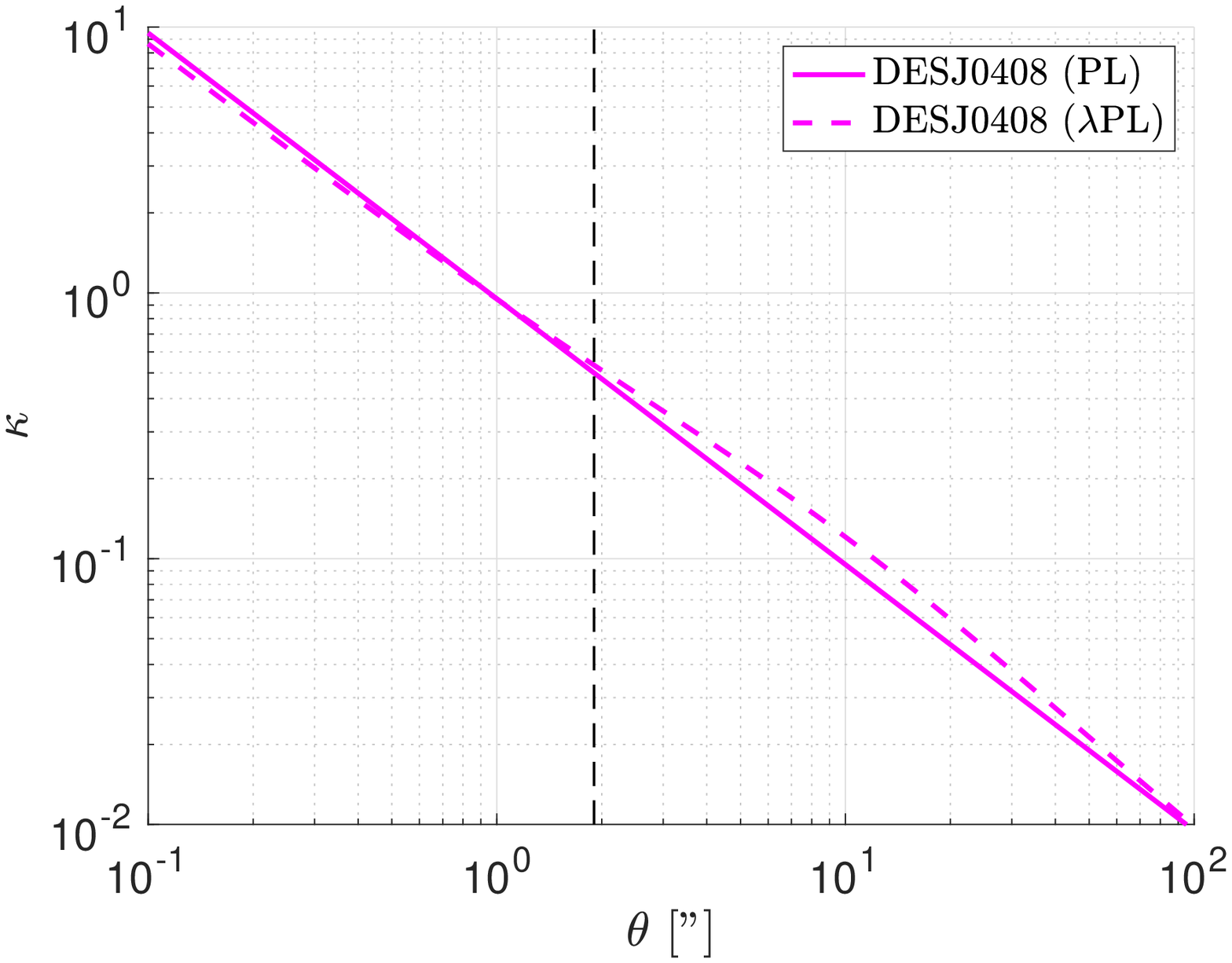}
    \includegraphics[scale=0.45]{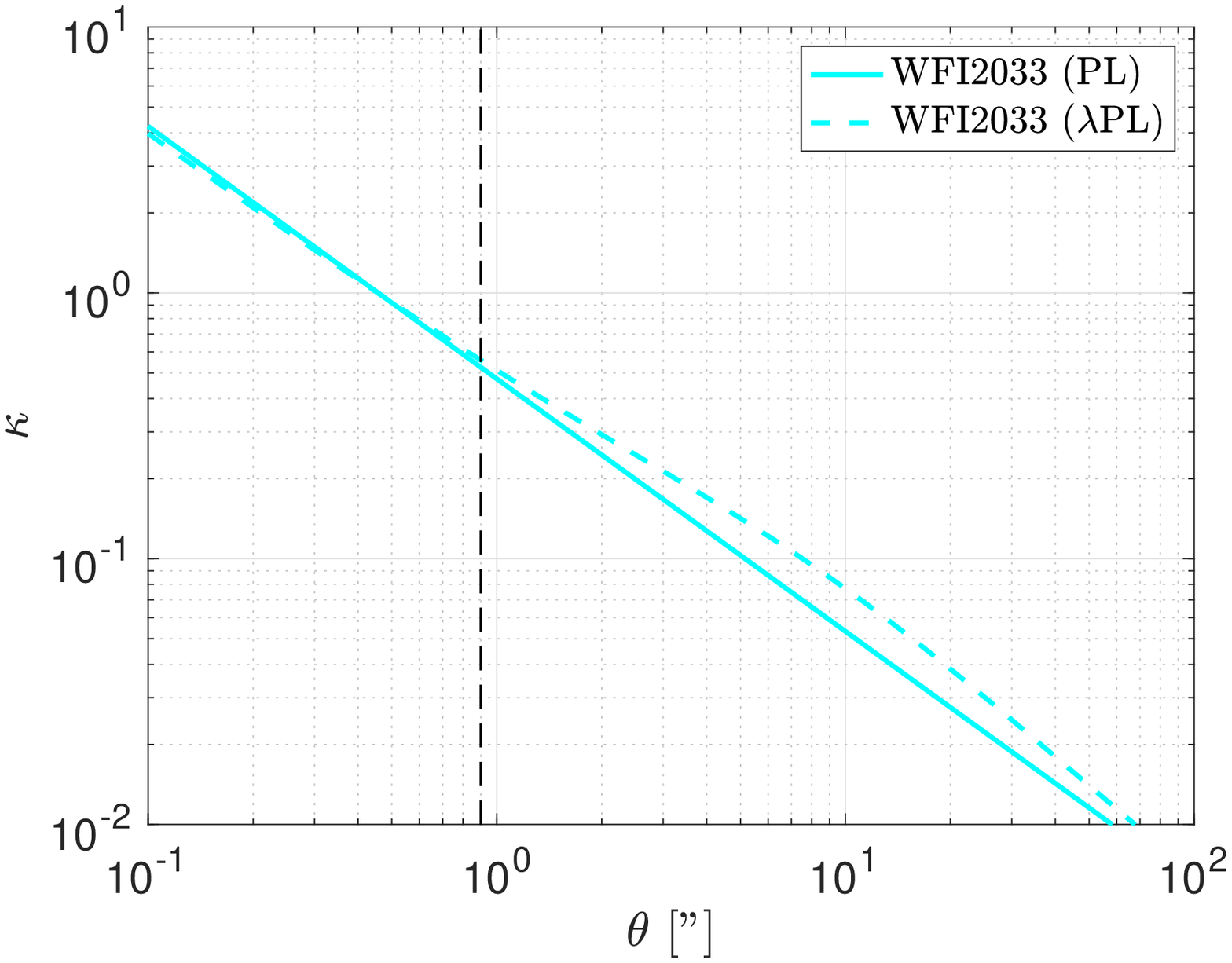}
 \caption{Inferred $\kappa$ profiles (1D projection) for the lenses in Tab.~\ref{Tab:lenses}. The vertical dashed line shows $\theta_E$ for each system. The core component is cNFW.}
 \label{fig:profiles}
\end{figure}

Noting that H0LiCOW found adequate fits to the lensing reconstruction with the PL model, and given an estimate of $\lambda$ for each system from Tab.~\ref{Tab:lenses}, we can use Eq.~(\ref{eq:PLlam}) with some model for $\kappa_c$ to investigate the implied physical shape of the lens galaxies.  In Fig.~\ref{fig:profiles} we show the results of this exercise for five systems\footnote{The 6th system -- J1206 \cite{Birrer19} -- has $\lambda=1\pm0.08$, so while it would admit a $\lambda\sim0.92$ core it is also consistent with no core component.}, where we use $\kappa_{\rm cNFW}$ with $\theta_s=11$'' and $\theta_c=5.5$'' to play the role of $\kappa_c$. For simplicity we ignore the ellipticity $q$ of the PL component. Including it would shift the PL line by a constant factor of $q^{\frac{\gamma-1}{2}}$ if we project along the $\theta_x$ direction, or $q^{-\frac{\gamma-1}{2}}$ if we project along $\theta_y$, without adjusting $\kappa_c$. Typical H0LiCOW lenses have $q\sim0.8$ and $\gamma\sim2$.
%


Considering Fig.~\ref{fig:profiles}, it is important to note that the particular shape of the profiles at angles $\theta>\theta_E$ is probably poorly constrained by the lensing data. Despite the fact that H0LiCOW utilises extended host information, in practice the host image pixels do not exceed $\theta\lesssim1.5\theta_E$ or so. As we have seen, the impact of the edge of the core component, which breaks the MSD, enters the image plane geometry at order $(1-\lambda)\theta^2/\theta_c^2$. The $\lambda$ values needed to sort out the $H_0$ tension imply $1-\lambda\approx0.1$ or so for most systems; even for the most deviant system (PG1115) we have $1-\lambda\approx0.2$. This means that $\theta_c\gtrsim3\theta_E$ would be enough to bring the MSD-breaking deformation down to the 1\% level for most systems. Moreover, in a real analysis, some of this deformation would probably be absorbed by the fitting for the source plane host parameters. A full-fledged analysis \`{a}-la H0LiCOW, fitting $\lambda$PL models to the real data, would be needed to truly quantify the level of the degeneracy. At this point, however, we emphasize that the shape of the profiles in Fig.~\ref{fig:profiles} at $\theta>\theta_E$ comes from our particular choice of $\kappa_c$ in this example, and is not necessitated by the data.


Finally, let us make a preliminary comparison with constraints from kinematics. \cite{Cappellari2015} presented an analysis of stellar kinematics in early-type galaxies with stellar masses in the range $\log_{10}\left(M_*/{\rm M_\odot}\right)\sim10.2-11.7$. These systems may be reasonable analogue systems to H0LiCOW lenses. According to~\cite{Cappellari2015}, the total density profiles of all of the analysed galaxies are consistent within the modelling uncertainties with simple PL all the way from $r\sim0.1R_e$ out to $r\sim4R_e$, where $R_e$ is the half-light radius. This range of kinematics coverage is interesting because it overlaps with and extends the range covered by the lensing analyses, which typically probe $r\lesssim R_e$.

\begin{figure}
\centering
 \includegraphics[scale=0.45]{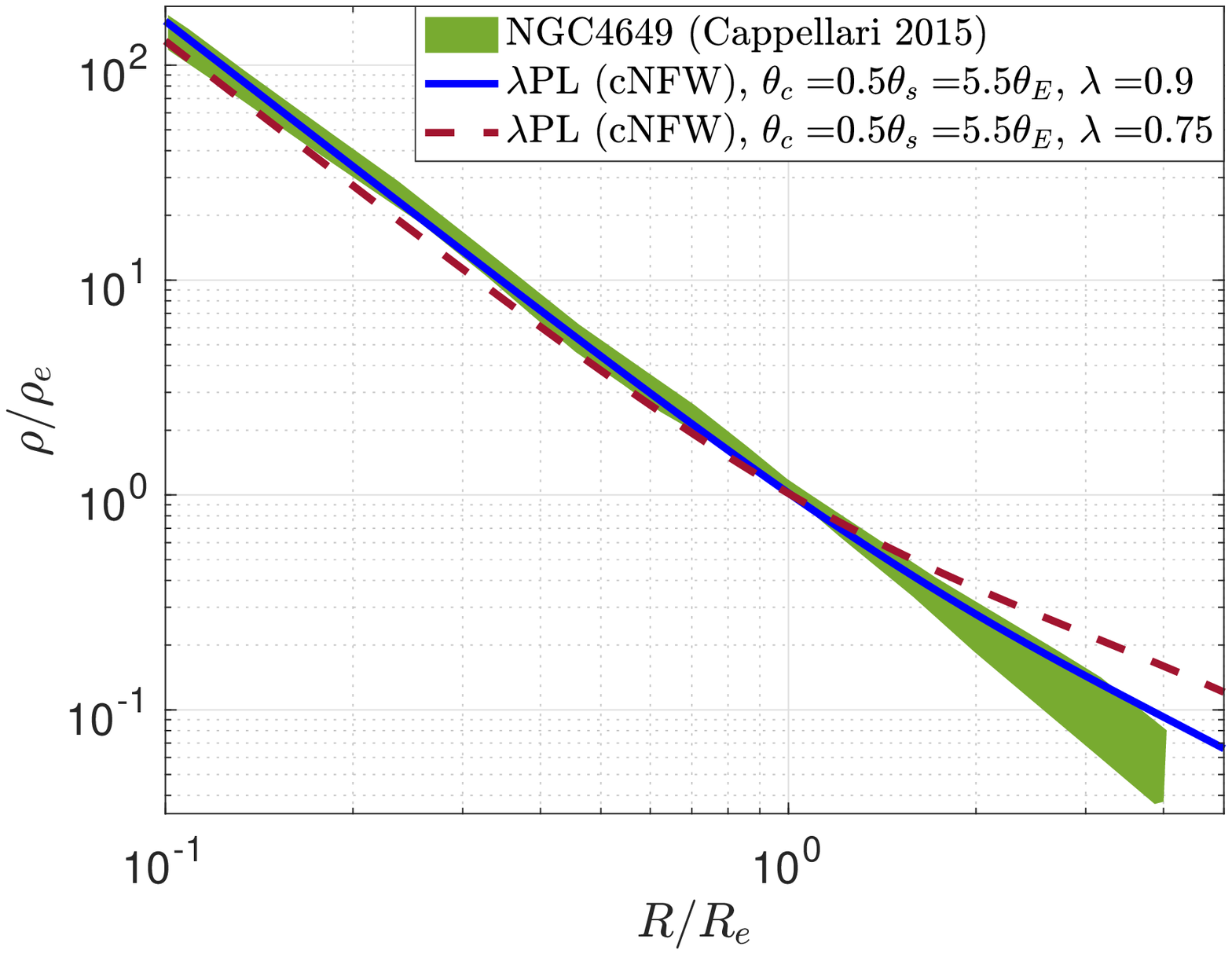}
 \includegraphics[scale=0.45]{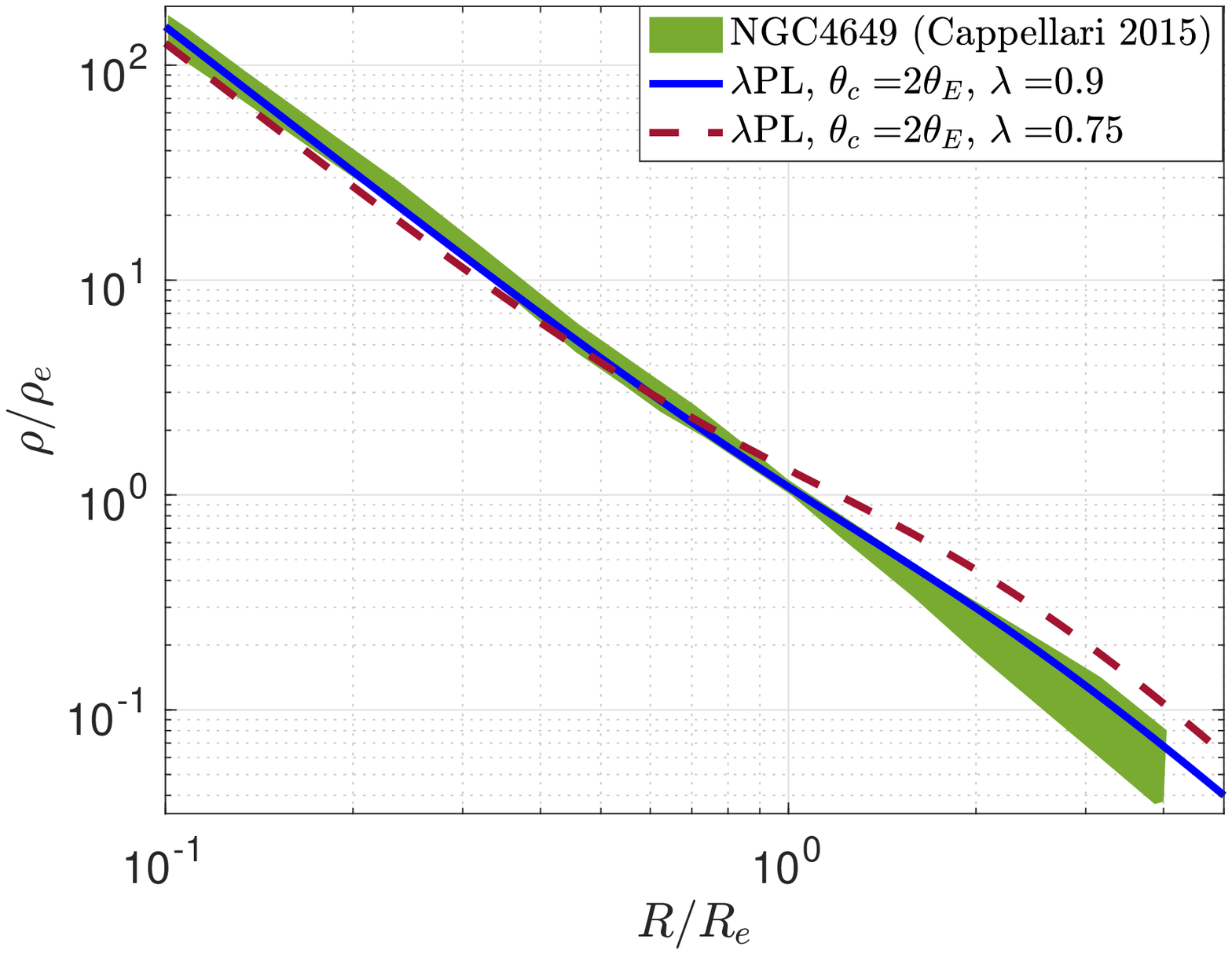}
 \caption{3D density: comparison with constraints from kinematics. Solid (dashed) line shows the $\lambda$PL profile for $\lambda=0.9$ ($\lambda=0.75$). Shaded band shows the posterior distribution of profiles from the kinematics fit of~\cite{Cappellari2015}. In this plot, for concreteness, we set $R_E=R_e$. 
 {\bf Left:} PL+cNFW. {\bf Right:} PL+cored PL.}
 \label{fig:kin}
\end{figure}

In Fig.~\ref{fig:kin} we compare the 3D density of a $\lambda$PL model with the profiles found in the galaxy kinematics analysis of~\cite{Cappellari2015}. The kinematics constraint, shown for the example of the system NGC4649 (see panel~d of Fig.~4 in~\cite{Cappellari2015}), is given by the shaded band that envelopes a collection of 100 profiles obtained by randomly selecting model parameters from the posterior distribution of the fit. The $\lambda$PL models for $\lambda=0.9$ and $\lambda=0.75$ are shown by solid and dashed lines, respectively. In the {\bf left} panel we show the 3D equivalent of the cNFW model considered in Figs.~\ref{fig:examplekappa} and~\ref{fig:example}. In this example, the PL component in the $\lambda$PL model is chosen to have\footnote{Note that~\cite{Cappellari2015} finds characteristic spectral index $\gamma>2$ for all of their halos, while the lensing analyses typically find a softer index $\gamma<2$.} $\gamma=2.25$.  
In the {\bf right} panel we show an example where the core component of the $\lambda$PL model is chosen to be a cored PL function $\rho_c\propto\left(R_c^2+r^2\right)^{-\frac{3}{2}}$ (see App.~\ref{app:1} for details). In both examples we assumed $\theta_E=\theta_e=R_e/D_l$.

The comparison of $\lambda$PL models to the results from~\cite{Cappellari2015} should be regarded with caution, as the family of dark matter density profiles considered in~\cite{Cappellari2015} was restricted to a generalised NFW form that does not overlap with the $\lambda$PL shape. With that in mind, we take Fig.~\ref{fig:kin} to suggest that currently, constraints from kinematics most likely cannot exclude $\lambda$PL with $\lambda\sim0.9$, which is the range of $\lambda$ that would be implied from cosmography if one calibrated $H_0$ from CMB/LSS data. This said, PL-core combinations with, e.g. $\lambda=0.75$ could perhaps be constrained by data, motivating a dedicated kinematics analysis specifically designed to test $\lambda$PL profiles.

\section{Discussion and Summary}\label{Sec:disc}
Lensing data alone cannot resolve the mass sheet degeneracy. Therefore, we think that the likelihood function in the cosmographic measurement of $H_0$ would have a very flat (albeit not completely flat) direction, corresponding to the effective MST $\lambda$ parameter of $\lambda$PL models. The H0LiCOW collaboration could thoroughly test this hypothesis on the real data. The $\lambda$PL models do not require more fitting parameters than, for example, the composite stars+NFW models considered already by H0LiCOW.

Stellar velocity data do resolve the MSD and could constrain $\lambda$PL models. Probably too large variations over the pure PL, in terms of the total enclosed mass, are not allowed by the stellar velocity data.
However, we doubt that stellar velocity data can test an $\lambda$PL model with $\lambda\approx0.9$. Either way, if the systematic uncertainty comes to be dominated by the stellar velocity modelling, then the significance of the $H_0$ tension would need to be revised. In this respect, we agree with the recent discussion of~\cite{Kochanek19}.

It would be very interesting, if indeed H0LiCOW has detected a core component in the lens galaxies. We are not aware of such cores in N-body or hydrodynamical simulations. Perhaps they could arise if, for example, the dark matter sector contains a component of ultralight (\cite{Schive:2014dra}) or self-interacting (\cite{Spergel:1999mh}) dark matter. On the other hand, since typical H0LiCOW lenses have $\theta_E\ll\theta_s$ (inferred in their composite stars+NFW models), it is clear that the lensing data probe the inner part of the lens halo where baryons either dominate the dynamics or at least make a large impact on it, making the simulations challenging. From this point of view, a detection of $\lambda$PL profiles with $\lambda\approx0.9$ could be turned into a goal to explain. Before we venture to more speculations, though, it would be reassuring to see a dedicated lensing/time delay analysis that tests $\lambda$PL models on the real data.

\acknowledgments
We are grateful to Michele Cappellari, Raphael Flauger, Claudio Grillo, Ranjan Laha, Sergey Sibiryakov, Yotam Soreq and, especially, Simon Birrer, Frederic Courbin, Aymeric Galan, Martin Millon, Alessandro Sonnenfeld and Sherry Suyu for useful discussions. KB is incumbent of the Dewey David Stone and Harry Levine career development chair.

\vspace{5mm}

\begin{appendix}
\section{Some examples of cored profiles}\label{app:1}
Here we first recall basic properties of the pure power law model, relevant for lensing analyses, and then give examples of cored profiles that can be used to test our $\kappa_c$ proposal. For simplicity we only consider isotropic models.

%
\paragraph{Pure power law.}
Consider the PL 3D isotropic density profile, designed to have projected Einstein radius $R_E$:
\be\rho_{\rm PL}(r)&=&\frac{\Sigma_c}{R_E}\frac{3-\gamma}{4\sqrt{\pi}}\frac{\Gamma\left(\frac{\gamma}{2}\right)}{\Gamma\left(\frac{\gamma-1}{2}\right)}\left(\frac{r}{R_E}\right)^{-\gamma}.\ee
The convergence, deflection angle, and potential for this model are:
\be
\kappa_{\rm PL}(\vec\theta)&=&\frac{3-\gamma}{2}\left(\frac{\theta}{\theta_E}\right)^{1-\gamma},\\
\vec\alpha_{\rm PL}(\vec\theta)&=&\left(\frac{\theta}{\theta_E}\right)^{1-\gamma}\vec\theta,\\
\psi_{\rm PL}(\vec\theta)&=&\frac{\theta^2}{3-\gamma}\left(\frac{\theta}{\theta_E}\right)^{1-\gamma}.
\ee
%

\paragraph{Cored power law.}
Consider
\be\rho_c(r)&=&\frac{\Gamma \left(\frac{\gamma }{2}\right)}{\sqrt{\pi }\Gamma \left(\frac{\gamma -1}{2}\right)}\frac{\Sigma_cR_c^{\gamma-1}}{\left(R_c^2+r^2\right)^{\frac{\gamma}{2}}}.\ee
The convergence, deflection angle, and potential for this model are\footnote{For $\gamma=3$ we have $\vec\alpha_c(\vec\theta)=\frac{\theta_c^2}{\theta^2}\ln\left(1+\frac{\theta^2}{\theta_c^2}\right)\vec\theta$ and $\psi_c(\vec\theta)=-\frac{1}{2}{\rm PolyLog}\left(2,-\frac{\theta^2}{\theta_c^2}\right)$.}:
\be\label{eq:kapgam}
\kappa_c(\vec\theta)&=&\left(1+\frac{\theta^2}{\theta_c^2}\right)^{\frac{1-\gamma}{2}},\\
\vec\alpha_c(\vec\theta)&=&\frac{2\theta_c^2}{\theta^2}\frac{\left(1+\frac{\theta ^2}{\theta_c^2}\right)^{\frac{3-\gamma }{2}}-1}{3-\gamma}\vec\theta,\\
\psi_c(\vec\theta)&=&\frac{2 \left(\left(1+\frac{\theta^2}{\theta_c^2}\right)^{\frac{3-\gamma}{2}}-1\right)}{(3-\gamma)^2}-\frac{H_{\frac{\gamma -3}{2}}+2 \ln \left(\frac{\theta}{\theta_c}\right) }{3-\gamma}+\frac{\theta_c^2}{2\theta^2}\left(1+\frac{\theta^2}{\theta_c^2}\right)^{\frac{3-\gamma }{2}} \frac{\Gamma \left(\frac{\gamma -3}{2}\right)}{\Gamma\left(\frac{\gamma+1}{2}\right)} \, _2F_1\left(1,1;\frac{\gamma +1}{2};-\frac{\theta_c^2}{\theta^2}\right).
\ee
Here, $H_\eta=\int_0^1dx\frac{1-x^\eta}{1-x}$ is the Harmonic Number and $_2F_1\left(a,b;c;x\right)$ is the Gauss hypergeometric function.

\paragraph{Cored NFW.}
Consider the cored NFW profile (cNFW) with core radius 
\be R_c&=&\zeta\,R_s,\ee
where $\zeta$ parametrizes the ratio between the core radius $R_c$ and the usual NFW scale radius $R_s$. To use this profile as a properly normalized core component, we write the 3D density as 
\be\label{eq:cNFW3D}\rho_{\rm cNFW}(r)&=&\frac{\left(1-\zeta\right)^2}{2\left(\zeta-1-\ln\zeta\right)}\frac{\Sigma_cR_s^2}{\left(\zeta R_s+r\right)\left(R_s+r\right)^2}.\ee
For this profile we were only able to find an analytical expression for the convergence,
\be
\kappa_{\rm cNFW}(\vec\theta)&=&\frac{1}{\zeta -1-\ln \zeta }\left(\frac{1-\zeta }{b^2-1}+\frac{2 \left(b^2 \zeta -1\right) {\rm ArcCot}\left(\sqrt{\frac{b+1}{b-1}}\right)}{\left(b^2-1\right)^{3/2}}-\frac{2 \zeta \, {\rm ArcCot}\left(\sqrt{\frac{b+\zeta }{b-\zeta }}\right)}{\sqrt{b^2-\zeta ^2}}\right),
\ee
where we define $b=\theta/\theta_s$. 
Near the origin, $b\ll1,\zeta $, we have 
\be \kappa_{\rm cNFW}&=&1+\frac{2\left(1-3\zeta ^2+2\zeta ^3\right)\ln \left(\frac{b}{2}\right)-2\ln \zeta +\zeta ^2 (4 \zeta -5)+1}{4 \zeta ^2 (\zeta -\ln \zeta -1)}b^2+\mathcal{O}(b^4),\ee
as required from our definition of a $\kappa_{\rm cNFW}$ component.

Because we consider a spherical profile, the deflection angle and the potential can be obtained by the following numerical integrals:
\be\label{eq:acnfw}
\vec\alpha_{\rm cNFW}(\vec\theta)&=&\frac{2\hat\theta}{\theta}\int_0^\theta dx\,x\,\kappa_{\rm cNFW}(x),\\
\psi_{\rm cNFW}(\vec\theta)&=&2\int_0^\theta \frac{dy}{y}\int_0^y dx\,x\,\kappa_{\rm cNFW}(x).
\ee

\end{appendix}

\clearpage

\bibliography{references}{}
\bibliographystyle{aasjournal}

\end{document}